# What-if Analysis for Business Professionals: Current Practices and Future Opportunities

Sneha Gathani
sgathani@umd.edu
University of Maryland, College Park
College Park, Maryland, USA

Zhicheng Liu
leozcliu@umd.edu
University of Maryland, College Park
College Park, Maryland, USA

Peter J. Haas
phaas@cs.umass.edu
University of Massachusetts, Amherst
Amherst, Massachusetts, USA

Çağatay Demiralp
cagatay@csail.mit.edu
AWS AI Labs & MIT CSAIL
Cambridge, Massachusetts, USA

## ABSTRACT

What-if analysis (WIA) is essential for data-driven decision-making, allowing users to assess how changes in variables impact outcomes and explore alternative scenarios. Existing WIA research primarily supports the workflows of data scientists and analysts, and largely overlooks business professionals who engage in WIA through non-technical means. To bridge this gap, we conduct a two-part user study with 22 business professionals across marketing, sales, product, and operations roles. The first study examines their existing WIA practices, tools, and challenges. Findings reveal that business professionals perform many WIA techniques independently using rudimentary tools due to various constraints. We then implement representative WIA techniques in a visual analytics prototype and use it as a probe to conduct a follow-up study evaluating business professionals' practical use of the techniques. Results show that these techniques improve decision-making efficiency and confidence while underscoring the need for better support in data preparation, risk assessment, and domain knowledge integration. Finally, we offer design recommendations to enhance future business analytics systems.

## CCS CONCEPTS

• **Human-centered computing → Empirical studies in HCI**; **User studies**; **Visual analytics**; *Graphical user interfaces.*

## KEYWORDS

Business Intelligence, What-if Analysis, Predictive and Prescriptive Analytics, Interview Study

## 1 INTRODUCTION

What-if analysis (WIA) is crucial for making data-driven decisions [20, 30]. For example, business managers may want to understand which marketing activities (e.g., sending emails, making phone calls, giving out free trials, running ad campaigns) should be prioritized to increase customer acquisition rates [30]; public health researchers might be interested in examining the effects of socio-economic factors (e.g., maximized smartphone use or low sleep quality) on adolescent obesity rates [49]; operations analysts may want to identify the optimal time complexities of various sub-algorithms used in a solver (e.g., cut generation, global constraint, local search) to reduce the solver's overall computation time [95]. To answer these questions, WIA is essential for identifying key data variables that influence specific outcomes, simulating various scenarios to understand how changes in data variables impact outcome predictions, and determining optimal data variable values to achieve desired target outcomes. For example, in marketing, WIA can be used to achieve a target outcome: increasing customer acquisition by 5%. The process begins by first identifying key variables such as ad campaigns and phone calls, and building models that predict customer acquisitions based on these variables. By exploring alternative scenarios with varying numbers numbers of campaigns and phone calls, businesses can assess their impact on acquisition rates. These insights enable data-driven decision-making to optimize marketing strategies effectively.

In such examples across many different domains, WIA plays an important role that traditional data analysis methods cannot fulfill. Similar to many analytics methods, WIA involves building models (e.g., statistical, probabilistic, neural network, optimization) that represents relationships between independent and dependent variables. However, unlike confirmatory methods such as hypothesis testing, WIA is inherently exploratory: it involves changing variable values in multiple scenarios and comparing possible outcomes without significance testing. On the other hand, exploratory data analysis (EDA) [93], which focuses on understanding trends and patterns in a dataset, is not appropriate for simulating alternative scenarios and predicting their outcomes to make data-driven decisions [20, 64]. As data continues to grow exponentially [21], limitations of human memory and cognitive capacity in generating hypotheses, testing scenarios, and predicting outcomes also exacerbate [30]. WIA thus addresses a critical need that is not supported by these methods, and a growing body of work is dedicated to developing WIA tools for various domains [25, 28, 31, 35, 77, 91, 98].

Despite efforts to support WIA, existing research mostly focuses on data analysts and scientists with technical expertise in coding, statistics, and algorithmic modeling. It overlooks domain professionals like healthcare experts, urban planners, or law officials [14, 20, 30, 64] who rely on no-code methods like graphical user interfaces and direct manipulation. These non-technical domain professionals frequently engage in WIA because they possess critical domain knowledge. For example, brainstorming alternative scenarios and evaluating potential trade-offs between variables [20] require extensive domain-specific experience and intuition—skills that professional data analysts may lack. Additionally, domain professionals play a key role in translating data-driven insights generated by data analysts and scientists into actionable strategies that align with their goals [10, 14]. Understanding their current practices and pain points is essential for developing better tools that effectively integrate domain knowledge with WIA techniques.

In this paper, we focus on a key group of domain professionals conducting WIA: *business professionals* [1]—such as sales managers, marketing managers, product managers, and operations managers. These professionals represent a significantly large and influential demographic group [74, 75] that regularly makes data-driven decisions without employing technical methods like coding and often has limited formal training in technical fields like computer science, machine learning, or statistics. They are also the primary audience for self-service business intelligence (BI) tools, which constitute a multi-billion dollar industry [37] and one of the most prominent commercial applications of visual data analytics [83]. Since little is known about how business professionals employ WIA and the challenges they face, we seek to answer the following research questions:

**RQ1:** What techniques and tools do business professionals employ to perform WIA for making data-driven decisions?

**RQ2:** How do they perceive and interact with advanced WIA techniques, and what future opportunities exist for improving them?

We answer these by conducting a two-part user study with 22 business professionals. For **RQ1**, we conduct a semi-structured interview study to understand their WIA workflows and challenges when making data-driven decisions. In addition to discussing their own experiences, we ask participants to walk us through a common business use case: maximizing sales by making spending decisions in various advertising channels, often referred to as marketing mix modeling. Our findings reveal that business professionals employ various WIA techniques but rely on rudimentary methods such as those provided by spreadsheet applications, which are insufficient for making data-driven decisions. Compared to prior studies on data workers' generic analysis practices (e.g., [20, 92]), our study uncovers a specific set of WIA techniques adopted by business professionals (e.g., sensitivity analysis, driver importance analysis, segmentation analysis) and the associated challenges they face. Moreover, business professionals prefer performing WIA independently, without relying on data analysts, due to (1) the limited availability of data analysts in enterprises, (2) the inefficiency of communicating with data analysts, (3) the business pressure to make quick decisions, and (4) the importance of incorporating domain knowledge, which data analysts often lack. These findings suggest that there is inadequate support for business professionals to perform WIA for making data-informed decisions.

The first interview study informs us of various WIA techniques employed by business professionals and the challenges they encounter with them. We address **RQ2** by first implementing four representative WIA techniques identified from the initial interview study (i.e., driver importance analysis, sensitivity analysis, goal-seeking analysis, and constrained analysis) in a visual analytics prototype. We then conduct a follow-up study with the same participants, asking them to use this prototype as a probe to make decisions for the same marketing mix modeling use case as in the interview study. Through the study, we identified the potential benefits and opportunities of WIA techniques for business professionals. Our findings show that they improve business professionals' decision-making speed and confidence, highlighting their strong interest in adopting additional advanced analytics techniques for informing their decisions. However, participants also expressed opportunities for more support in preparing data (e.g., consolidating data from multiple sources, consistently defining KPI goals), assessing prediction risks (e.g., understanding confidence and trust in predictions), and incorporating domain knowledge (e.g., capturing limited budgets, accounting for volatile ecological and market conditions such as COVID and inflation).

In summary, this paper contributes:

- An **interview study** identifying WIA techniques, tools, and challenges that business professionals face when making data-driven decisions.
- A **task-based study** gathering business professionals' hands-on experience and feedback on the benefits and opportunities of using and enhancing WIA techniques for their use in making data-driven decisions.
- **Design recommendations** for future business analytics systems to better support business professionals in advanced analytics.

## 2 RELATED WORK

Our work relates to prior research on other professionals performing data analysis, use of enterprise dashboards, advanced analytics tools, and organizational decision-making.

### 2.1 Understanding Professionals Performing Data Analysis

Prior work has examined how different types of professionals perform data analysis. Table 1 summarizes the data, high-level tasks, and levels of technical engagement (i.e., low, medium, high) in some related works examining data analysis across various audience roles, providing context for the landscape.

Many studies focus on professionals using technical approaches, like data analysts, data scientists, and business analysts, who mainly support decisions rather than directly make them. For example, Kandel et al. [45] identify challenges in analysts' data pipelines, while Alspaugh et al. [1] explore challenges in exploratory analysis. Demiralp et al. [18] study data scientists' exploratory analysis, and Zhang et al. investigate business analysts' needs in generating reports [101] and narratives [100]. Newburger and Elmquist study statisticians' decision-making focusing on visualization [25]. Similar to business professionals, these people work with multivariate data consisting of both categorical and quantitative attributes. However, they have strong technical backgrounds and engage in exploratory or confirmatory data analysis, which is distinct from WIA in decision-making.

[1] In the remainder of the paper, we use the terms *business professionals* and *professionals* interchangeably.

| Paper | Job Role | Data | High-Level Task | Technical Engagement |
|---|---|---|---|---|
| Kandel et al. [45] | Data Analysts | Multivariate (categorical and quantitative) | Exploratory data analysis, build statistical models | High: programming expertise |
| Alspaugh et al. [1] | Data Analysts | Multivariate (categorical and quantitative) | Exploratory data analysis | High: programming and GUI tools |
| Demiralp et al. [18] | Data Scientists | Multivariate (categorical and quantitative) | Exploratory data analysis | High: Python programmers |
| Zhang et al. [101] | Business Analysts | Multivariate (categorical and quantitative) | Making reports and dashboards | Medium: limited ML/data science backgrounds |
| Newburger and Elmquist [25] | Statisticians | Multivariate (categorical and quantitative) | Exploratory and confirmatory analysis | High: scripting to build statistical models |
| Bartram et al. [6] | Data Workers | Multivariate (categorical and quantitative) | Sensemaking | Low: no ML/data science backgrounds |
| South et al. [84] | Political Scientists | Text | Exploratory analysis and sensemaking | Low: no ML/data science backgrounds |
| Jasim et al. [39] | Community Leaders | Text | Analysis and decision-making | Low: no ML/data science backgrounds |
| Kang and Stasko [46] | Investigative Analysts | Text and documents | Sensemaking | Medium: limited ML/data science backgrounds |
| Bhattacharya et al. [10] | Healthcare Experts | Multivariate (categorical and quantitative) | Exploratory model steering | Low: no ML/data science backgrounds |
| Khadpe et al. [48] | Line Managers | Nominal/qualitative list, value tree | Decision-making | Low: no ML/data science backgrounds |
| **Our Work** | **Business Professionals** | Multivariate (categorical and quantitative) | **Improve outcomes and decision-making** | **Low; no ML/data science backgrounds** |

**Table 1: Landscape of data, high-level task, and technical engagement of some related research works on data analysis by target audience in a variety of job roles. Our research focuses on *business professionals*, who are domain experts involved in improving and making decisions using multivariate data but do not engage in technical tasks like programming.**

Bartram et al. [6] study "data workers" who engage in data analysis as part of their daily tasks but lack formal training in technical domains like computer science, statistics, or machine learning. Previous work often focuses on data workers in non-business domains, such as political science [84], community leadership [39], and law enforcement [46]. These data workers are similar to business professionals in terms of their low technical expertise. However, many of them work on data exploration and sense-making tasks [50, 51, 69, 76] instead of decision-making, and do not discuss processes that involve simulating alternative scenarios and action selection [12, 13, 57, 64]. For those data workers who need to make decisions, they primarily work with unstructured data like text and nominal data [22, 39, 48], which present different challenges from multivariate data. By studying business professionals who perform WIA, our work focuses on a unique combination of the data, task, and technical level dimensions as shown in Table 1.

## 2.2 Use of Dashboards in Enterprises

Previous work investigates the role of visualization and dashboards in enterprise data analysis and decision-making. For example, studies find that business professionals, like sales managers, frequently use dashboards [7, 63, 96]. However, they are used mainly to communicate findings to stakeholders, not to make data-driven decisions. Dimara and Stasko [19] note the lack of studies on decision-making in visualization research, and other works [20] show how visualizations can be integrated into long-term decisions. Yet, business professionals face challenges in creating dashboards and conducting analysis necessary for making decisions [79, 92]. This is due to incomplete information displayed in dashboards and their limited interactivity, which restricts professionals from drilling into the data to explore different scenarios or conduct simulations. Therefore, visualization alone without data-driven analytics is insufficient, leading us to focus on how advanced analytics can support or hinder decision-making.

## 2.3 Advanced Analytics Tools

Marx et al. [59] and Bergeron et al. [8] studied legacy information systems in organizations or Executive Information Systems (EIS), identifying gaps in decision-making support, especially in drill-down, scenario analysis, and optimization. Van et al. [94] explored systems for marketing management, echoing these limitations. More recent work by Crisan and Correll [14] and Bartram et al. [6] also noted these shortcomings, though they focused less on business decisions. Zhang et al. [100] and Honeycutt et al. [36] emphasized integrating domain knowledge with AI to enhance user trust in intelligent systems. Bhattacharya et al. [10] empower healthcare workers to utilize AI model explanations to improve model predictions. Oral et al. [64] highlight the lack of decision support features like the visibility of alternatives, input processing, change model parameters, etc. in existing tools. Our study expands on exploring model-based WIA techniques for making data-driven decisions, highlighting challenges and opportunities for integrating domain-specific expertise into them.

Most human-in-the-loop machine learning research centers on data scientists and analysts, highlighting their openness to automating decisions [97] or collaborating with models [70]. Crisan and Fiore-Gartland [15] discuss automation's role in routine tasks, rapid prototyping, and democratizing data science, but for data scientists. Dimara et al. [20] share similar decision-making practices of decision makers. Our study differs by focusing on how specific group of users; business professionals who perform analysis using non-technical means perceive advanced WIA, a form of AutoML used for data-driven decisions. Further, we go beyond interviews, collecting empirical data from users *actively engaging* in *hands-on* decision-making with these techniques.

Many business intelligence (BI) tools [27, 80, 88, 89] and spreadsheet applications offer some advanced analytics, but they primarily focus on descriptive and exploratory tasks, answering the "what" and "why" in data. However, they often lack predictive and prescriptive features, leaving the question of "now what?" unanswered, making it difficult to plan subsequent actions. Tools like Excel's `SOLVER` [60] and `GOAL SEEK` [61] provide optimization features for desired outputs [6], but manual formula creation remains challenging for business professionals, which interactive advanced analytics tools aim to address.

## 2.4 Organizational Decision-Making

Management and organizational theory categorize decisions into three tiers: strategic (executive level), tactical (middle management), and operational (front-line employees) [5]. This work focuses on tactical decisions, which are short-to-medium term actionable steps (e.g., introduce discounts, run campaigns, reduce shipping costs, etc.) aimed at achieving specific business goals (e.g., compete with competitors' price changes, decrease churn rate, increase throughput, etc.). Additional examples are provided in Section 5.1. These decisions are dynamic, continuously evolving, and need improvement as they are central to the responsibilities of business professionals.

The three-step model by Simon [82] (also observed in [20] and used in [64])—intelligence (information gathered and problems identified), design (possible solutions for problems generated and evaluated), and choice (best alternative selected)—describes the decision-making process. Berisha-Shaqiri provides a more detailed eight-step model [9]. On similar lines, Jun [40] centers on hypothesis formalization through statistical analyses by data scientists. We observe similar steps in our participants' decision-making, but we focus on the data they use, the advanced analytics they apply, and the challenges they face in selecting the best alternatives. Perkins et al. [68] examine how marketing managers' experience influences their decisions and the information they use, while Little [54] reports on processes followed by marketing managers from 40 years ago, but with the aid of descriptive and exploratory tools. In all these works, the role of data and analysis is largely simplified or overlooked. In contrast, our study focuses on how managers employ data and advanced analytics across varied departments.

# 3 OVERVIEW OF THE TWO STUDIES

Here, we outline the methods and procedures for our studies.

## 3.1 Participants

We recruited 22 business professionals via an online survey [3] through the User Interviews [38] website and company's mailing lists. Our survey ensured that participants were selected based on three criteria: (1) *use* organizational, customer, or product *data* to guide business decisions, (2) *make* decisions *at least weekly*, and (3) have *no formal training* in data science or programming. The final participant pool comprised of business professionals from four departments: marketing (7), sales (5), product (5), and operations (5) managers from varied company sizes, sectors, and expertise levels, as shown in Table 1 in supplementary materials.

## 3.2 Protocol

Each participant completed both parts of the study in 120–140 minutes in two sessions, held on separate days within one to two weeks, as illustrated in Figure 1. All interviews were conducted, recorded, and transcribed online for analysis. Participants received a total of $175 in Amazon gift card for completing both sessions. Each study is detailed in later sections.

## 3.3 Analysis

We analyzed the recorded and transcribed data from both parts of the study using iterative coding to derive our findings. The first author, hereafter referred to as the coder, primarily conducted the coding. The coder double-checked with the other co-authors whenever questions arose or clarifications were needed to ensure rigor and consistency. For the first interview study, the coder began by conducting an initial broad categorization on participants' business decisions, data utilization strategies, and general WIA techniques and tools. This included identifying recurring patterns in how participants approached making data-driven decisions and integrated WIA into their workflows. In the second iteration, the coder focused on understanding the usage of specific WIA techniques (e.g., scenario exploration and ranking correlations between drivers and key performance indicators (KPIs)) and associated challenges (e.g., current tools being simplistic, needing manual efforts). In the third iteration, results from the first two rounds were then synthesized to form the overall findings of the interview study, providing a comprehensive understanding of participants' WIA workflows and challenges when making data-driven decisions.

For the follow-up task-based study, the coder analyzed participants' experiences of practically using four WIA techniques identified from the previous interview study and implemented within a visual analytics prototype probe. Over subsequent iterations, the analysis was revised to categorize the use of the functionalities, immediate reactions, challenges, and feedback. The coder then categorized these findings further, grouping feedback into potential benefits (Section 7.2) and opportunities (Section 7.3).

In both studies, the coder consulted with the other authors over subsequent iterations to ensure validity. This collaborative process ensured that the categories were coherent and accurately reflected the data. Ambiguities, such as conflicting interpretations of a participant's comments, were resolved through discussions with all authors. We support our findings with representative participant quotes throughout the paper.

# 4 STUDY I: INTERVIEW STUDY

Here, we present the method of the first interview study.

## 4.1 Method

The interview study was conducted through semi-structured interviews with business professionals to understand various WIA techniques they employed, motivations behind their use, tools they relied on, and challenges they encountered. It is outlined in Figure 1 and described underneath.

*4.1.1 Protocol.* Participants satisfying our selection criteria were interviewed remotely for 60-70 minutes, with one researcher facilitating and recording. Each interview had two parts: a general session followed by a task-based session.

**General Session.** The general session involved 21 open-ended questions to establish the broader context about their decision-making. These open-ended questions enabled participants to share their existing workflows in-depth and flexibly using their own use cases, offering insights across four categories: (1) role and team, (2) business goals, (3) data, tools, and WIA techniques, and (4) ideal tools, as detailed in Table 2. This session lasted 45 minutes.

**Task-Based Session.** Participants then spent 15-25 minutes on a task-based session discussing WIA techniques, tools, and challenges for a specific business use case of Marketing Mix Modeling (MMM). Discussion on this specific use case allowed for consistent comparisons across participants.

**Marketing Mix Modeling Use Case.** Participants, asked to imagine themselves as marketing managers, were tasked with making decisions to achieve the business goal of maximizing TV sales for the next quarter by assessing a historical dataset. The dataset consisted of weekly data on investments made on five advertising channels (SMS, TV, Internet, Radio, Newspaper), economic factors (Demand, Supply, Unit Price, Consumer Confidence Index, Producer Price Index, Consumer Price Index, Gross Rating Points or reach of advertising channels), and company's sales.

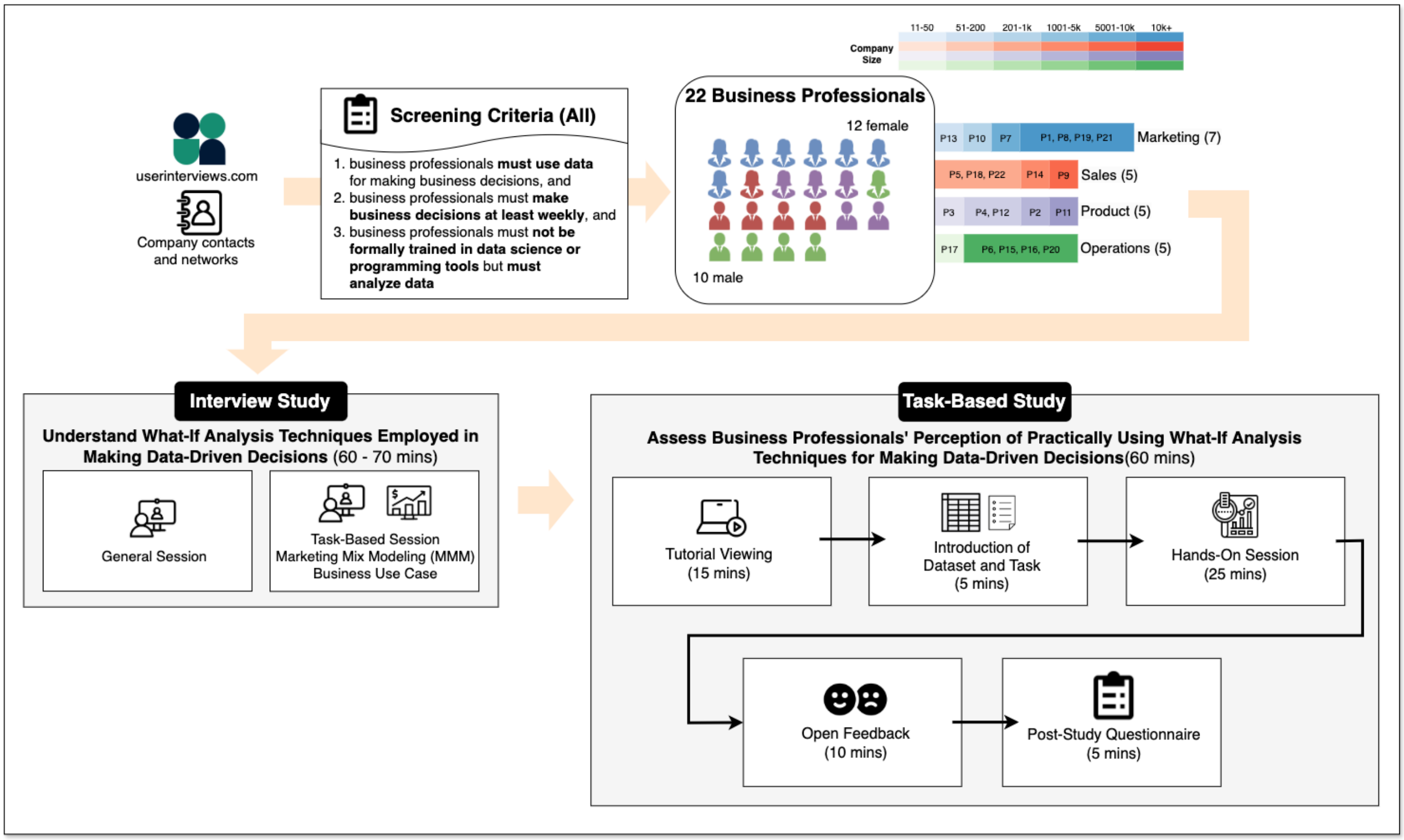

**Figure 1: Overview of our two-part study. We conduct the study with 22 business professionals (12 female, 10 male) from four departments: marketing (7), sales (5), product (5), and operations (5). The interview study aims to understand various what-if analysis (WIA) techniques participants employ, the tools they use, and the challenges they face to use them for their data-driven decision-making of both their own business use cases as well as for a specific business use case of Marketing Mix Modeling (MMM). The task-based study aims to understand the same participant's perception of practically using representative WIA techniques identified in the first study and implemented as a probe within a visual data analytics prototype. Colors refer to the departments in which the participant worked ( marketing , sales , product , operations ).**

Participants were shown this dataset, sourced from Kaggle [44] in a CSV format during task briefing. And prompted to discuss advanced analysis techniques and tools they would employ to analyze the data and make decisions to maximize sales. This use case was chosen since it was relatable for many, especially marketing managers, and broad enough for non-marketing participants to apply their own experiences to as well. Assumptions and validity of both this use case and dataset were verified by marketing experts at a company, who served as our collaborators. Participants could take notes.

# 5 FINDINGS OF THE INTERVIEW STUDY

We present the interview study findings in three parts: participants' decision-making context (goals, decisions, and data used), the WIA techniques and tools employed along with specific challenges encountered, and the broader challenges with advanced analytics.

## 5.1 Business Professionals' Goals and Decisions

We provide context on participants' business goals, decision questions, and data used.

*5.1.1 Goals.* Participants, regardless of their role or company, shared a common goal: improving business outcomes and increasing business efficiency to gain a competitive edge over competitors. These outcomes were measured through *Key Performance Indicators* (KPIs), such as sales, deal closing rate, and retention rate. Decisions focused on maximizing two predominant KPIs—the *revenue* and *sales*—while optimizing time and resource constraints. Given the complexity of pursuing these goals, they were broken into smaller, department-specific KPIs. For example, marketing managers P1 and P8 needed to make decisions for the goal of turning prospective customers into real ones, which is measured by the KPI *conversion rate*. In another example, product managers P4 and P12 were interested in preventing existing customers from leaving, which is measured by the KPI *retention rate*.

*5.1.2 Decisions.* To achieve their goals, business professionals made decisions about actions related to *drivers*, which were data variables about customers and products. Examples included search engine optimization (SEO) data (e.g., traffic, impressions), product usage data (e.g., feature usage, time spent), and customer activities (e.g., sign-ups, clicks). For instance, to boost conversion rates, decisions included "providing 3 months free trial on the product on signing up for a demo" (which involved the "demo sign up" driver), or "sending out targeted marketing emails showing how the product works for prospective customers' use cases" (which involved the "marketing emails sent" driver). Similarly, to increase retention rate, a plausible decision was "changing placements of product features and highlighting certain features using pop-ups that match customers in a certain geography" (which involved the "feature usage frequency" driver), or "providing tutorial use cases of product mapped to customers' business niche" (which involved the "customer vertical" driver). More examples are included in Table 2 in supplementary materials. For a given goal, participants needed to decide both *which* drivers to act on and *how* to modify them.

| # | Interview Study Questionnaire |
|---|---|
| **Role and Team** | |
| 1 | Briefly describe your role in your company. How many people are on your team, and what are their roles? |
| 2 | Are there data analysts in your team? |
| **Business Goals** | |
| 3 | How often does your daily work require you to answer questions or make decisions to achieve your business objectives? |
| 4 | Describe recent business questions that you/your team often try to answer. |
| 5 | Approximately how many hours or days do you spend answering each of these questions? |
| 6 | Do you answer these questions in one analysis sitting or repeatedly come back to find answers that would eventually help make a decision for your team or company? |
| **Data, Tools, and WIA Techniques** | |
| 7 | How often do you/your team use the data to make decisions? |
| 8 | Describe the data you or your team utilize in your work. |
| 9 | What are the tools or systems (could be commercial, your company products, or self-developed) you generally use to answer such questions? |
| 10 | Where does this data come from? |
| 11 | Talk us through the process of coming up with strategies or making decisions, elaborating on analytic techniques you use in response to one or two specific business questions you shared. |
| 12 | Do you consider your current process of using analytics you use efficient? Explain. |
| 13 | Do the set of tools you use for performing your analytics seem efficient? |
| 14 | What are the challenges you face to achieve your objectives? |
| 15 | What are the challenges of the set of tools you use? |
| 16 | Do you use any advanced analytics like making predictions using the data for your decision-making? Why/Why not? |
| 17 | Do you base your predictions on your intuition? Explain. |
| 18 | Do you base your predictions on your experience? Explain. |
| 19 | Have you ever heard of *WIA*, *predictive analytics*, and *prescriptive analytics*? Can you describe some examples? |
| 20 | Do you use any of these analytics that use models or similar techniques (e.g., machine learning, deep learning, statistical modeling, techniques, etc.)? If yes, which ones, why, which tools, expertise level, number of years of experience, examples of how. If no, why not and which techniques would you like to use? If sometimes, why? |
| **Ideal Tool** | |
| 21 | If you were to invent the perfect tool to work with data and make decisions with, what would that tool look like? What would it do? |

**Table 2: Questions that we asked participants during the general session of the interview study. These questions aimed to understand the context around business professionals' roles and teams, business goals, data, tools, WIA practices, and desired tools for making data-driven decisions.**

## 5.2 WIA Techniques Employed by Business Professionals

Participants used various advanced analytics techniques to understand how independent driver variables relate to dependent KPI business goals and to form hypotheses about these driver variables.

*5.2.1 Tasks.* They followed the following set of tasks to understand this relationship:

**Frame Questions.** With KPI goals in mind (e.g., maximizing sales, optimizing media mix), participants framed decision questions like *"what is the relation between the drivers and the KPI?" (P22)*, *"what penetrates" (P15)* or *"is driving" (P17, P20)* the sales, and what to do to *"diversify spending [across channels] to minimize diminishing ROI" (P3)*. Many of these questions aimed at exploring potential scenarios through WIA.

**Formulate Hypotheses.** Participants often had preconceived hypotheses about how drivers impacted KPI goals, based on their experience, expertise, and external factors. For instance, several participants (P1, P3, P4) shared P8's expectation that they *"...expect[ed] newspaper ads to be less effective than internet"*.

**Translate Hypotheses into Models.** Participants refined their hypotheses into simple quantitative models. For example, a marketer (P19) aimed to optimize impression rate (i.e., number of customers reached or tapped by a marketing action) by understanding how marketing channels (e.g., ads, website) contributed to the impression rate KPI by weighing how much each channel was contributing towards one impression. They came up with a rough hypothesis that if a potential customer looked at two paid media ads and visited the website, it would result in a complete impression, which was translated into a simple linear model (e.g., 2 × paid media ads + 1 × website visit = 1 impression.

**Revise Models with Data.** Participants frequently adjusted their models by varying parameters based on intuition and trial-and-error. Data played an essential part, and participants relied on existing business intelligence (BI) reports, visualizations, and raw data to refine their models. For example, after reviewing more data and playing with more models, P19 updated their formula to include video completion rate to lead to a more successful impression, changing the model formula to 2 × ads + 1 × website visit + 1 × video completion rate = 1 impression instead. This highlights the need for business professionals to test various driver-action scenarios [20, 97], similar to the learnings of [20, 97] studies, and observe its impacts on a given KPI before making strategic decisions.

**Explore Multiple Scenarios through WIA Techniques.** All participants explored multiple what-if scenarios to examine outcomes for various hypotheses and eventually make data-driven decisions. The participants described both WIA techniques that were already employed in their existing workflows, as well as techniques they wanted to use. For example, nearly all participants (20 of 22) manipulated driver or KPI values to observe their effects on other variables in the model in their existing tools, while the remaining participants expressed interest in using such advanced analytics. Figure 2 shows the WIA techniques in the order of popularity among the participants. Below, we describe each of the techniques and the current tools (or lack thereof) used by participants.

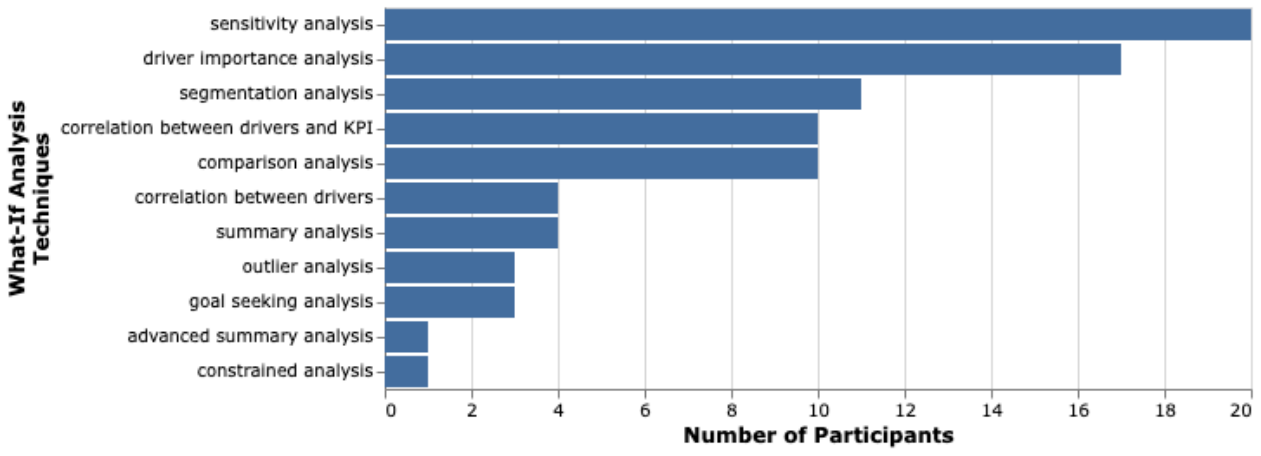

**Figure 2: What-if analysis techniques shared by the participants for informing their decisions using data.**

- *Sensitivity Analysis:* A large number of participants, 20 out of 22, wanted to observe how changes in driver values (e.g., advertising spend across different channels) affect key performance indicators (KPIs) like sales. For example, P7 wanted to know if *"they should do something to the unit price so the return [sales] is better?"* Such analysis is commonly referred to as *sensitivity analysis*, where professionals explore how altering input variables influences output variables [73].
  All 20 participants performed this using spreadsheets. They created models or formulas for calculating KPIs, as described in earlier tasks and manually adjusted values of various driver columns to observe the recalculated KPI formulas to avail this technique. They also tweaked formulas to simulate and experiment with different scenarios to explore the impact of changes on various driver combinations, helping guide decisions on the next steps. Although they performed this technique, participants complained about manually needing to track the change in their KPIs when alterations to the driver values were made. While a few participants (2 out of 22) mentioned existing tools like Google Analytics, Power BI, and Salesforce that offer text-based insights or narratives of scenarios, they felt these tools were too static. They preferred more dynamic, interactive tools that allow for real-time manipulation of drivers, enabling them to test multiple scenarios and see the immediate impact on KPIs.
- *Driver Importance Analysis:* 17 out of 22 participants desired to determine which drivers (e.g., advertising channels) were most significantly influencing the sales KPI, often known as *key influencers* or *driver importance analysis*. Currently, they all relied on an indirect way of learning driver importance by manually formulating hypotheses about how drivers related to sales KPIs and performing sensitivity analysis.
  Participants typically used trial-and-error, to edit the driver values from far negative to far positive values and observe corresponding formula calculations of sales that has the greatest impact on sales. This manual table editing approach relied heavily on domain knowledge and self-constructed formulas for KPIs, making it difficult to fully understand driver-KPI relationships. Participants acknowledged this limitation, as their approach relied heavily on domain knowledge and guesswork. Although tools like Salesforce Einstein offer a feature that ranks drivers by importance, none of the participants mentioned using it. One participant (P8) noted seeing a similar feature in Power BI's waterfall chart, which is also seen in Einstein, but found it hard to interpret when quick, actionable insights were needed.
- *Segmentation Analysis:* Half the participants performed *segmentation analysis*, that involved understanding the needs and behaviors of various drilled-down groups of data such as different "regions" or "user types", as shared by P10, P14, P18. Participants used this technique to tailor their strategies for distinct data segments. To do this, sorting and filtering techniques were employed to retrieve relevant data, but participants manually segmented data in spreadsheets, duplicating slices and dices of data into separate worksheets for individual analysis and comparison.
  No existing tools were mentioned as being directly helpful in this context. While platforms like Tableau and Salesforce Einstein provide text-based insights on certain segments, participants noted them to be limited in scope, not iterable or controllable [29], and often one-off that may not always be actionable or relevant to their specific needs. Additionally, if business requirements for a segment change, these could not be reflected in the data underneath to simulate its impact effectively.
- *Correlation Analysis and Outlier Analysis:* 10 of 22 participants expressed the need for learning *correlations between different drivers and their KPI* to help identify whether, in which direction, and to what extent changes in the drivers are associated with changes in the KPI. To make it easier to digest, they also wanted to learn a *summary* of the impacts.
  Of 10 participants, only three to four participants had conducted *correlation analysis between drivers* to learn the trend of a range of changes made to the drivers and their impact on the KPI. This was performed by creating heatmaps over tables to visualize correlations. However, participants expressed uncertainty about how to justify or communicate the implications of shifting results to other team members when model formulas were changed.
  Additionally, about three to four participants also mentioned detecting anomalies in the data using *outlier analysis*, where they currently relied on manual inspections (e.g., checking for values that are out of bounds or deviate significantly from expected ranges) for detecting outliers.
- *Comparison Analysis:* Half the number of participants also desired the abilities to perform *comparison analysis* where they could actively compare two or more of any of the advanced analyses, but found this to be difficult and confusing currently. Currently, participants needed to manually duplicate previous tables or analyses to compare them with changes made, resulting in a tedious process and a higher risk of human error.
- *Goal Seeking Analysis:* Few participants came up with a set of potential driver actions to trigger the desired changes in their KPI goals. P7, for example shared *"...send to different email, ...change subject line, ...colors, ... time"* actions to run an email promotional campaign to get *"1000 more downloads"*. This was currently based off of their experience and domain knowledge, but aligns with the basic concept of *goal-seeking analysis*.
  Only two participants referenced to an existing tool feature that enables this analysis: Excel's `GOAL SEEK` macro, which predicts the necessary driver values to meet a target KPI. However, one participant (P11) commented on the required high level of expertise to effectively use it, while the other participant (P21) mentioned not using *"it enough or have the time enough to use it... but can see the value..."* if the feature was interactive and easy.

- *Constrained Analysis:* Participants also explicitly shared budget and resource constraints that they took into account when making decisions. For instance, a participant P2 shared they *"obviously have a limited budget about what campaigns to do..."* and wanted to know if there were any features or tools to include these constraints during analysis, rather than relying on manual data adjustments. This is an example of *constrained analysis*.

**Formulate Initial Actions.** After exploring multiple scenarios, participants devised actionable decisions based on their analysis. For example, P19's revised model led to doubling resources on paid media ads and instructing sales managers to push prospective customers to visit the website and see the complete video explicitly.

*5.2.2 Tools.* All participants mostly used spreadsheet applications, primarily Excel, for the WIA techniques. In Excel, business professionals relied on manual methods to perform various WIA analyses described above, such as explicitly creating formulas and running calculations to explore multiple hypotheses, editing driver columns and experimenting with formula calculations to analyze sensitivity and driver importance, and using operations like filter and pivot to manually segment raw data for segmentation analysis. They also duplicated data for comparisons. In contrast, as reported from related work, technical professionals employed programming tools and machine learning models for their analyses [1, 15, 45]. Only one participant, P10, reported using advanced functions and features (e.g., LOOKUP, VLOOKUP to retrieve data, GOAL SEEK) but highlighted a lack of time to set up and use them.

Participants also created charts, but only to observe patterns (e.g., seasonal trends), behaviors (e.g., sales best in the west region), or anomalies (e.g., sudden sales drop every 5 years). Only 5 of 22 visualized their data in dashboards (static or interactive) using a variety of BI tools like Power BI, Salesforce, and Tableau, while 17 others relied on creating Excel charts. Those with access to technical data teams (8 out of 22) or other third-party application tools (e.g., Google Analytics, Marketo, and Pendo) were able to generate charts regularly. However, those who could not get customized reports easily often struggled with using the BI tools on their own.

To make data-driven decisions, most participants (17 out of 22) mentioned performing complex analyses described earlier (Section 5.2.1), however executing them heavily using manual and simple processes like updating tables and utilizing formulas. The descriptions provided by other participants implied a similar need for advanced analytics and collectively all participants called for more sophisticated means to conduct them easily.

*5.2.3 Challenges.* We identified several broader challenges faced by business professionals in leveraging advanced analytics for making data-driven decisions, detailed as follows:

**Advanced Analytics Using Rudimentary Methods.** Even though a wide range of existing analytic tools are available, business professionals resorted to using rudimentary methods such as manual table editing, trial-and-error approaches to create formulas, and manual copying and duplicating tables. The participants explained the reliance on these rudimentary methods due to several reasons. First, challenges with advanced analytics are amplified for business professionals due to their lack of formal technical training. While spreadsheets were acknowledged as *"super powerful" (P11)*, the participants often lacked the expertise or *"experience with [more complex] macros, etc. which not everyone has[d]" (P11)*. As a result, they had to use basic or outdated tool features, making even simple analyses inefficient, time-consuming, and prone to errors. For example, P11 recounted frustration dealing with extremely large boilerplate Excel sheets made 20 years ago by some individual and how *"absolutely non-replicable the process is and how the person who developed it is still contacted at times to understand some niches"*. Such workflows hinder collaboration, reproducibility, and scalability, especially for analyzing larger datasets. Secondly, many state-of-the-art BI tools offer limited prescriptive or predictive analytics capabilities. Participants appreciated the descriptive and exploratory features in current BI tools. For example, the visual data exploration capabilities of Tableau were praised, calling them *"spectacular" (P5)* and *"Excel on steroids!" (P4)*. However, these tools offer only basic data statistics, percentage of goals achieved, and limited drill-down scenarios, further highlighting the rudimentary nature of these descriptive methods. Finally, even when a tool offers advanced analytic features, its usability and learnability often pose a challenge. Participants noted that many analysis features were *"clunky..., ...not very user-friendly..." (P5)*, and they *"need[ed] to have developer support to use it... and extensive training" (P4)*. An employee at one of the BI tool companies also admitted that they *"are not proud of not using [their own tool], but it does have a steep learning curve" (P21)*. Participants also highlighted the lack of state-saving features, limiting their usefulness for iterative analyses. Collectively, these issues illustrate why the participants shied away from advanced features and employed rudimentary methods.

**Business Professionals Conduct Advanced Analytics without Relying on Technical Professionals.** All participants reported a shortage of analysts in companies, regardless of their size, to handle business professionals' requests. They mentioned that data teams also have limited bandwidth and prioritize company-wide initiatives (e.g., *"developing models for churn rate analysis" (P4, P5)*, *"customer ordering patterns" (P6)*, *"staff hiring models" (P5, P6)*) and automation over their needs of analyses queries. Even when analysts were available, participants expressed frustration due to analysts' lack of domain knowledge and inefficient communication which made the collaboration *"a lot of added work [inconvenience] in itself" (P3)*. Additionally, under the pressure to make quick decisions, business professionals often did not have the time or patience to engage with data analysts or scientists, but spent exorbitant hours, mostly 1-4 hours (20 out of 22) and at times 4-6 hours (2 out of 22) daily on analysis themselves without relying on them. These findings highlight a gap between ideal collaborative practices and the real-world challenges of achieving effective communication and understanding and goal alignment between business professionals and data analysts [17, 71].

**Unique Challenges Exist Beyond Data Analysis.** Before delving into advanced analytics to understand driver-KPI relationships, business professionals often perform tasks similar to those of data analysts [1, 45], such as acquiring, cleaning, and enriching data. However, they face greater challenges of integrating data from various sources, and ensuring data quality, as they lack technical expertise. When they conduct advanced analytics independently, without the support of technical data scientists or analysts, they face additional challenges. For example, they must balance various business factors, such as budget constraints, economic principles (e.g., the impact of price increase on sales), market conditions (e.g., COVID, interest rates), and social considerations (e.g., employee work-life balance), all while under pressure to make accurate and actionable decisions. Also, participants perform analyses continuously, where they experiment with hypothetical scenarios, refine hypotheses, formulate decision actions, share them with stakeholders, and conduct A/B testing over smaller groups of customers over multiple iterations to execute decision actions that eventually help achieve their goals. Consequently, business professionals' analyses to make data-driven decisions are inherently more complex, extending beyond traditional exploratory data analysis.

# 6 STUDY II: FOLLOW-UP TASK-BASED STUDY

The first interview study informed us of the workflows business professionals use to make data-driven decisions, the various WIA techniques they employ, how they currently execute them, the difficulties they face, and the techniques they desire to have. Building on these findings, in the task-based study, we wanted to learn how the participants perceived and interacted with advanced analytics techniques. To achieve this, we implemented four representative WIA techniques identified in the previous study within an interactive visual analytics prototype. This served as a probe to gather participants' hands-on experience and feedback on using WIA techniques for making data-driven decisions. This approach aimed to address the challenges and limitations of participants' existing workflows and tools by providing a more intuitive and streamlined experience.

## 6.1 Method

We outline the methodology of our task-based study, followed by a description of the WIA techniques implemented in a visual analytics prototype to use as a probe.

### 6.1.1 *Protocol.*

We conducted a 60-minute remote session with the same participants for the task-based study with the following protocol and as shown in Figure 1.

**Tutorial Viewing.** We provided a 15-minute pre-recorded video of the visual analytics prototype to acquaint participants with the implemented WIA techniques. The demo used another common business case, *deal closing analysis*, which sales managers use to increase customer acquisition rate by analyzing prospective customer's interaction data with the product and the organization.

**Re-Introduction to the MMM Business Use Case and Task.** Participants were reintroduced to the MMM use case and dataset to establish the context for the hands-on session. They were then briefed on the same task of maximizing TV sales, but this time with instructions to apply representative WIA techniques using the prototype. This lasted 5 minutes and participants could take notes if desired.

**Hands-On Session.** Participants accessed the prototype via an AWS-hosted application link. Based on a pilot study with 3 participants (not included in the 22), we gave each participant 25 minutes and broke down the task into 10 manageable sub-tasks, as tabulated in Table 3). They were instructed to think aloud, given a document detailing each sub-task, and received assistance as needed. Participants were additionally prompted to share how each technique could be applied to their own decision-making use cases. The coder took notes on interesting participant interactions (e.g., switching between different functionalities), clarify questions asked, or feedback on using the functionalities (e.g., goal-seeking analysis needed multiple iterations to understand).

**Open Feedback.** Following the hands-on session, participants provided 10 minutes of feedback on the WIA techniques' utility, benefits, concerns, and any additional analytics they wanted. This session also gathered insights on their trust, confidence, and perceived effectiveness of the techniques. The coder also took notes of the feedback to ensure that any missed feedback from the open-feedback questionnaire (Table 3) was collected.

**Post-Study Questionnaire.** In the final 5 minutes, participants completed a post-study questionnaire [4]. This questionnaire included Likert scale-based questions to capture participants' thoughts on the usefulness and usability of the functionalities, as well as open questions to gather quantitative ranking of functionalities based on usefulness, concerns with functionalities, and any additional functionalities they would like. Details are in the supplementary materials [2], and all sessions were recorded for further analysis.

## 6.2 Representative WIA Techniques Implemented into a Visual Analytics Prototype to Use as a Probe

We chose four WIA techniques from the seven techniques identified in the previous study and implemented them within an interactive visual analytics prototype to use as a probe in this task-based study. We did not implement all seven techniques to avoid overwhelming the participants. The four techniques were driver importance analysis, sensitivity analysis, goal-seeking analysis, and constraint analysis. They were considered representative because they either directly encompass other WIA techniques or serve as important building blocks for more complex WIA techniques. For example, segmentation analysis, which involves filtering and drilling down into specific data groups, can be viewed as a preliminary step to applying techniques like driver importance or sensitivity analysis on data subsets. Similarly, correlation analysis, which focuses on understanding relationships between drivers or between drivers and KPI, is inherently part of driver importance analysis, as it seeks to identify and quantify the impact of drivers on KPI and between each other. Outlier analysis can also be integrated within these techniques by first identifying anomalies and then applying sensitivity or constraint analysis to understand and manage their impact.

Participants were informed that the visual analytics prototype was only a probe to capture professionals' practical and critical feedback, not an exhaustive set or tool. Throughout the study, participants were also encouraged to consider other advanced analytics techniques that could further enhance their decision-making. We also informed the participants that the techniques were applicable to other use cases beyond the MMM scenario, like increasing customer retention, decreasing churn rate, and optimizing inventory.

| # | Task-Based Study Sub-Tasks and Open-Feedback Questionnaire |
|---|---|
| **Sub-Tasks** | |
| 1 | Using driver importance analysis technique, what are the top three and least three drivers that influence the sales KPI? |
| 2 | Using sensitivity analysis technique, which advertising channels will you invest in in order of most to least to increase the sales KPI? |
| 3 | Using the sensitivity analysis technique, what is the sales achieved if you increase the Demand driver by 5% and the Supply driver by 3%? What can you say about the demand-supply relationship? |
| 4 | Using the sensitivity analysis technique, find the percentage increase in the Unit Price driver needed to achieve the same or more sales if the Demand and Supply drivers don't change. |
| 5 | Using the summary analysis technique, discuss the trend observed of perturbing the non-advertising channel expenses drivers between a range of -100% to 700% and step size of 100% on the sales? Do you expect this behavior? |
| 6 | Using the summary analysis technique, discuss the trend observed of perturbing the advertising channel expenses. Which advertising channels would you invest or not invest in to increase the sales of the TV? Do you expect this behavior? |
| 7 | Using the goal seeking technique, what is the maximum TV sales you can achieve with the original boundary conditions? |
| 8 | What is the sales uplift that you can achieve if you only optimize the advertising channel drivers and the Unit Price driver? |
| 9 | How much will the sales increase if you have constraints of $2-$115 on the Unit Price driver? |
| 10 | Can you achieve a target sales of $980 million? If yes, what is the driver that needs the maximum uplift? Do you think this is feasible? |
| **Open-Feedback Questionnaire** | |
| 1 | Were these specific tasks relatable to the ones you would try to answer to achieve your goal of increasing sales or any other decisions that you take in your own work? |
| 2 | Do you understand the key driving advertising channels behind sales? |
| 3 | Do you understand the relationship between the advertising channel expenses and sales? |
| 4 | How open would you be to using such WIA techniques for answering your own business questions? |
| 5 | What would be some other advanced analytics techniques that would be helpful for you to achieve your goal? |
| 6 | Were you able to reach a decision on investments needed to achieve maximum sales? Explain. |
| 7 | How confident are you in the decision you reached? Explain your reasoning. |
| 8 | What are your concerns about your inferred decision? Explain your reasoning. |
| 9 | What were the challenges you faced to achieve the goal of increasing the sales of the TV company? |
| 10 | Do you consider the analysis process you currently used of using the WIA efficient? Explain your reasoning. |

**Table 3: In the hands-on session of the task-based study, we asked participants to complete a set of sub-tasks (bottom) requiring them to use what-if analysis techniques implemented in an interactive visual analytics prototype. In the subsequent open feedback session, we asked participants to answer a set of questions (bottom) in a questionnaire, eliciting their qualitative feedback based on their task completion experience.**

We described the representative WIA techniques in an order reflecting how business professionals typically approach understanding driver-KPI relationships. Findings from our study revealed that participants often preferred to use them in any order and iteratively.

**Driver Importance Analysis.** This technique helped understand the relative importance of drivers in predicting the sales KPI, such as the top three advertising channels to prioritize. We implemented this technique as an ordered list of drivers (Figure 3B), accompanied by a bar chart (Figure 3A) showing the computed importance values, offering a simple to read, systematic, and insightful approach compared to approaches seen in existing tools. Using this, professionals quickly learned that the top and bottom three drivers influencing the sales KPI were CCI, PPI, and Supply, and Newspaper Ad Expenses, GRP of Radio, and GRP of Newspaper, respectively. Additionally, participants could interactively select a subset of drivers they were interested in to assess the importance of those specific drivers in relation to the KPI.

**Sensitivity Analysis.** This technique helped observe the effects of adjusting driver values, such as increasing or decreasing expenses on different channels, on the sales KPI. This was provided in the form of coordinated views, comprising a perturbations pane (Figure 3C), a bar chart (Figure 3D), and an uplift pane (Figure 3E). Participants could input a numeric or percentage change to adjust one or more drivers, which automatically updated the bar chart and uplift panel to reflect the corresponding KPI output prediction. The average sales observed on the original data was 96 million dollars (blue bar); upon perturbing the "supply" driver by 3% and "demand" driver by 5%, the predicted sales dynamically shows an increase by 3.63% (equivalent to 3.5 million dollars) (green texts), making the overall predicted sales 100 million dollars (yellow bar).

Summary analysis (Figure 3G) allowed professionals to view multiple sensitivity scenarios at once, aiding them in comprehending the sales KPI trend for when individual driver values underwent perturbation across a specified range of bounds (Figure 3F). The perturbation range could be added as numeric values or percentage change. Small multiples chart was used, with each multiple displaying the sales KPI trend when each advertising channel expense was varied between -100% (lower bound) to 500% (upper bound), with a step size of 100%, revealing significant sales increases for SMS and Radio channels, a slight rise for the Internet channel, and declines for TV and Newspaper channels. Participants could hover over the small multiples to view a tooltip displaying the original KPI value alongside the predicted KPI value after the perturbation.

**Goal-Seeking Analysis.** The technique helped professionals learn predictions of the required driver values to achieve an optimum or user-desired aggregate KPI value. For example, they can directly identify the actions needed to maximize the sales KPI. We implement it interactively (Figure 3H), where participants chose their optimization goal (maximize, minimize, or set a target goal), observed the pre-specified constraints on all drivers to initially be optimized between 0% to 300%, and ran the analysis. On maximizing the sales KPI goal, a prediction of 549.07 million dollars sales could be achieved (470.11% uplift) along with the driver values shown in a table having average original, predicted, and change values that enabled this prediction.

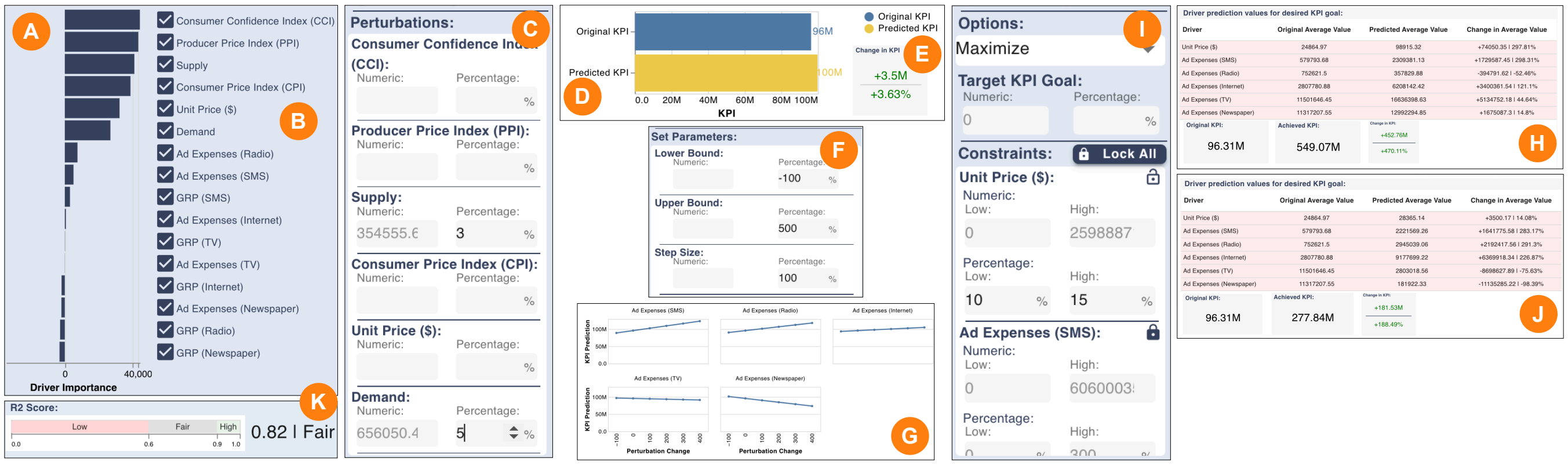

**Figure 3: Interface showing the four representative WIA techniques implemented in a visual analytics prototype, used as a probe to capture participants' practical use of WIA techniques for making decisions in the MMM business use case during the task-based study: (1) driver importance analysis bar chart (A) with toggle-able drivers (B), (2) sensitivity analysis perturbation pane (C), bar chart (D), and uplift pane (E), along with its variant of summary analysis (G) with bounding pant (F), (3) goal-seeking analysis prediction (H), (4) constrained analysis options and constraints pane (I), and prediction (J), and the *R*2 accuracy score pane (K).**

**Constrained Analysis.** This technique incorporated user-defined business constraints (e.g., boundary, equality or inequality conditions) on one or more drivers and run goal-seeking analysis to get driver values satisfying these constraints. For instance, when using the previous goal-seeking analysis technique, participants observed that in order to maximize sales, the "unit price" driver needed to increase by 297.81% or 74000 dollars, which customers would not practically pay. In response, they could restrict the "unit price" driver to increase between only 10% to 15% by specifying it in the constraints pane (Figure 3I). Maximizing sales again reduced the sales prediction from 549.07 million to 277.84 million dollars, but ensured constraining the "unit price" driver within the specified limits to 14.08%, yet showing a significant uplift of 188.49% (Figure 3J).

The performance of the underlying prediction model was presented, either as accuracy (measured by the $R^2$ score) or error rate. For example, the MMM use case demonstrated a fair performance with an $R^2$ score of 0.82 (Figure 3K).

# 7 FINDINGS OF THE FOLLOW-UP TASK-BASED STUDY

We present our findings by first reporting the participants' experiences of using the representative WIA techniques for the MMM business use case. Then, we discuss the potential benefits and opportunities for enhancement that were identified for the use of advanced analytics for making data-driven decisions.

## 7.1 Insights From Using Representative WIA Techniques for the MMM Business Use Case

Feedback from both the open-feedback session and post-study questionnaire showed unanimous agreement that WIA techniques implemented in the visual analytics prototype probe were *"helpful" (P2, P10, P12, P17, P20)*, *"interesting" (P3, P8)*, *"powerful" (P13, P16)*, and quick to *"set [business professionals] in the right direction" (P7, P17)* for making decisions to maximize the sales KPI. All participants comfortably completed all 10 sub-tasks within the allotted time, finding them simple and easy to use, and made similar decisions, such as increasing the unit price of TVs, boosting SMS advertising spending, and cutting newspaper ad spending. They also mentioned that using WIA techniques was *"better than sitting looking at the data" (P9)* that they did in their existing workflows. All but one participant were excited to apply the techniques to their own business use cases, to *"play around with own data [since it would be] super helpful when setting goals for the quarter" (P3)* and learn data behaviors like *"how it can relate to their business" (P6)* and *"how it would be useful for drilled-down segments [subsets of their data]" (P4)*. The remainder participant, an operations manager, found the techniques to be *"okay and wanted [them] to be more dummy-proof" (P15)* suggesting their utility, but with the need for greater customization towards their domain.

Participants collectively acknowledged the usefulness of all the techniques. Figure 4 illustrates the relative rankings of the techniques. Sensitivity analysis was rated the most useful, followed by constrained analysis, goal-seeking analysis, and driver importance analysis. The goal-seeking technique required the most guidance (6 out of 22 participants), despite aligning well with their decision-making needs, highlighting the importance for explicit guidance, demonstrations, and examples for their practical use. While participants also initially struggled with interpreting the small multiples chart expressing summary analysis, they found it valuable for reducing cognitive load in experimenting with multiple what-if scenarios. However, they stressed the need for simpler visuals when presenting such advanced analyses to executive stakeholders.

Further, participants expressed a keen interest in additional advanced analytics techniques that they also mentioned in the first interview study. They sought automated features with robust drill-down capabilities, enabling segmentation analysis to help answer *"how different geographical locations are doing?" (P14)*, *"realize how the data behaves for different verticals of customers?" (P18)*, or *"how*

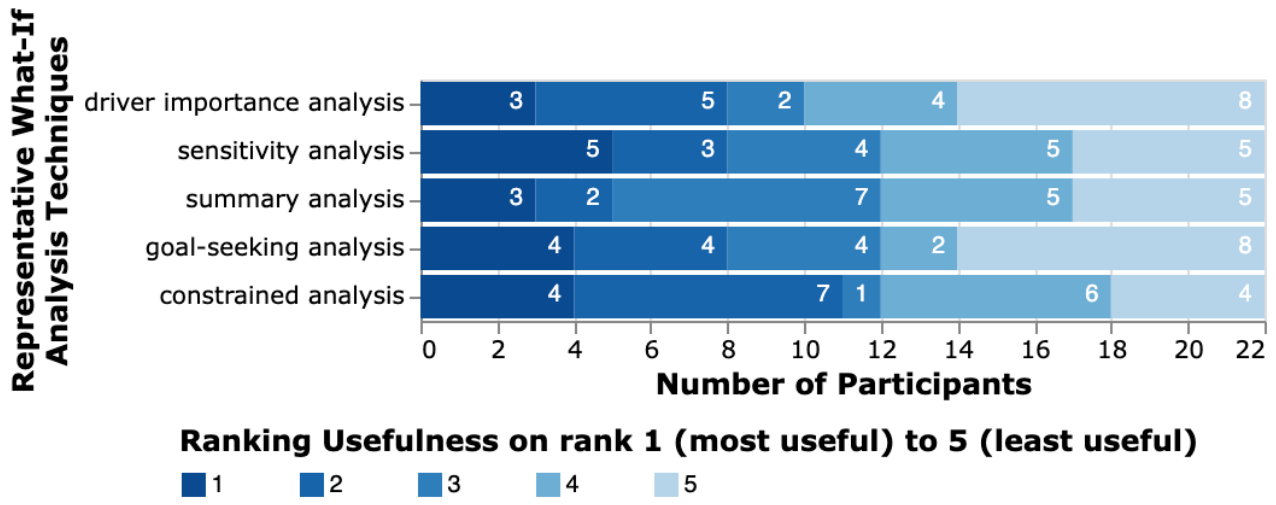

**Figure 4: Ratings of the what-if analysis techniques from most to least useful. Participants found sensitivity analysis to be the most useful technique.**

*to leverage existing customers and products doing well for those not performing?" (P21)* They also wanted techniques to uncover correlations between drivers and KPIs and facilitate easier scenario comparisons. Participants requested enhancements to currently provided WIA techniques too, like seamlessly integrating results from one technique into another (e.g., plugging in the predictions from goal-seeking analysis back into sensitivity analysis (P21)) for continuing the analysis, and the ability to save analysis results for comparisons (P1, P8, P14).

## 7.2 Benefits of Using WIA in Making Data-Driven Decisions

We report three key benefits shared by participants in using WIA techniques for making their data-driven decisions.

*7.2.1 Helps Accelerate Decision-Making.* Participants valued the ability of WIA to provide a *"holistic view" (P13)* of driver-KPI relationships, a stark contrast to the current manual, spreadsheet-based methods they currently rely on. More importantly, they appreciated how WIA enabled them to quickly observe predictions across multiple scenarios, incorporating various driver combinations-all without requiring external expertise from data analysts or specialized teams. Consequently, all participants agreed that WIA would accelerate their decision-making and *"increase [their] efficiency" (P6)*. Many were also pleasantly surprised by the capabilities of the representative techniques, noting that while they had encountered *"something similar, [it was] not anywhere close to these" (P11)*. In previous experiences, they often had to customize or build their own tools to achieve similar functionality. This underscored participants' recognition of WIA's transformative potential in streamlining and enhancing decision-making through advanced analytics.

*7.2.2 Increases Confidence in Decisions.* Participants articulated that WIA helped them move away from guesswork and trial-and-error approaches, thereby bolstering their confidence in their decision-making [23], as it narrows down the vast possibilities of hypotheses to those observed in the data. One participant P1, for instance, happily shared how WIA can *"help understand where to put more effort!"* Moreover, participants emphasized that leveraging advanced analytics techniques, particularly interactively as demonstrated in our study, could help them think and test out a lot more corner cases and scenarios that they might miss in traditional manual analysis. Further, participants shared how they could play with many more drivers to observe their consequences on the business KPI and be able to understand the market, product, and customer behavior more widely and better. All these reasons combined led the participants to excitedly share how they can build a strong narrative or story to support their decision actions. A participant P10, for example, expressed how WIA will enable *"everyone in the company [to] be on the same page about a decision... no one's decision will be overwritten"*. Therefore, participants showed promise in WIA enabling them to be confident in their decisions when communicating with the stakeholders.

*7.2.3 Streamlines Decision-Making and Enhances Teamwork.* Almost all participants agreed that their existing analysis processes could not be replicated easily to justify their decision actions. For instance, multiple participants commented that their current decision actions were based on *"trial and error" (P3, P14, P19, P21)*, which they often deemed to be inefficient and unpredictable, making it challenging to rely on. Recognizing the limitations of their current practices, participants saw the potential of WIA to streamline decision processes, making them more accessible, reproducible, and actionable. This consequently allowed for *"consistent" (P3, P10)* [45] and *"smoother" (P22)* decision-making. Further, in addition to WIA techniques helping them in their own analysis, participants also added that it will help facilitate collaborative analyses for decision-making envisioning a scenario where stakeholders could engage in decision reasoning collectively. For example, one participant P4 mentioned how they could *"take [this] to their job today without the support of IT... also give it to other team members"*. Therefore, participants explained that advanced analytics techniques would help them make robust decisions as they could get external perspectives and benefits from others' experiences and expertise through standardized and collaborative analysis processes.

## 7.3 Opportunities in Enhancing the Use of WIA for Making Data-Driven Decisions

After learning the benefits of WIA, we outline three opportunities raised by our participants to enhance the utilization of WIA techniques for making data-driven decisions.

*7.3.1 Currently Inaccessible to Business Professionals.* Many participants like P1, P7, P8 mentioned that even though they had heard of features similar to some representative WIA techniques in existing commercial BI tools (e.g., key influencers in Power BI, importance and correlation analysis in Einstein, etc.), they had *"no idea or not know too much [about them] or barely used them"*. Upon further discussion, participants revealed that the high costs associated with such tools rendered them inaccessible. Additionally, participants voiced concerns about the complexity of these tools, which typically offered a plethora of models and parameters to navigate, or deemed them to be too general and inefficient to use for their own use cases. The overwhelming number of features too made it difficult for professionals to effectively use the tools, a challenge exacerbated by automatically generated results that lacked clear explanations for them to draw meaningful conclusions. Moreover, participants cautioned the requirement of technical expertise to utilize these tools effectively, a skill set that many business professionals did not possess. Furthermore, business professionals also lacked the time and incentives to learn or to keep up with such technologies, which are rapidly evolving.

Despite their enthusiasm for adopting WIA in their work, it seems that tools are not currently designed to meet their needs. The vast space of possible scenarios seems to be a major problem in particular. P15, for instance, frustratingly mentioned how they have *"...to [currently] commit to just solving the first thing... that is going to be the most bang for the buck.... [because they]... have 900 problems happening all at once..."*. Such expressions inform us that business professionals have no time to experiment with all possible scenarios but need flexibility to quickly identify the first set of actionable decisions that yield favorable business outcomes and play with them to account for constraints, a concept akin to "satisficing" [82], suggesting that more complex analyses atop WIA, like recommendations or ranking of WIA scenarios, could help.

*7.3.2 Difficulties in Interpreting and Understanding Predictions.* While participants expressed enthusiasm about using WIA techniques for decision-making, they emphasized that analytics alone cannot make the decisions. Their domain expertise remains critical for developing effective strategies [97]. For instance, P4 pointed to how *"common human understanding may not translate to the model without [their] external involvement"*, giving an example where an increase in unit price beyond a threshold will decrease demand, thereby decreasing sales rather than increasing it, a nuance that may not be fully captured by models. Participants emphasized the need for transparency in how predictions were generated and wanted clarification on *when and to what extent* to adopt predictions. For instance, they wanted to know how to handle *"unrealistic [predictions]" (P5, P6, P22)*, or when *"so many factors [or decision paths]" (P2)* are shown to them, and how to *"[include] factors that cannot be captured in the data" (P2, P18)*.

Another recurring challenge was interpreting the accuracy of predictions (low v/s fair v/s high) and translating this into confidence in next steps [23]. For example, should low-scored models be treated as worst-case scenarios and perform additional manual analysis to make decisions? Or should they take medium scored models as baseline and expect risks in lower scored models? While senior, more experienced business professionals understood that predictions should be taken with caution, they emphasized the importance of training junior professionals not to follow predictions blindly. Addressing these concerns requires WIA tools to foster deeper transparency and provide interpretable predictions that can help enhance business professionals' trust in the predictions.

*7.3.3 Lack of Support for Data Preparation and External Constraints.* Participants reported various challenges in data gathering, consolidation, featurization (e.g., curating new hypothesized drivers for KPI goals), aggregation data from multiple sources, and cleaning it to work around missing, inaccurate, and outlier data. These tasks are already complex for expert analysts [1, 15, 20, 45], and our study confirms that it can be even more daunting for business professionals. For example, the participants highlighted issues such as incorporating qualitative data (e.g., meetings, chats, and reviews) as drivers into models, handling interdependent drivers (e.g., calls and meetings), and resolving ambiguities related to the drivers or KPIs, which often carry different meanings across teams. To address these issues, participants suggested co-designing WIA techniques with seamless integration into data wrangling and preparation systems.

Although participants valued techniques like goal-seeking and constrained analysis to incorporate their domain expertise, they worried about the mechanisms *"on setting the right constraints, coming up with them, how to set them right and optimize on that" (P5)*. On a related note, some participants expressed concerns over not being aware of ideal conditions for the team and company, thereby gauging the constraints either conservatively or over-confidently. For example, P5 mentioned only knowing that they *"had five million dollars for advertising... [and] can set only overall constraint over all channels"* making it challenging to set constraints on each of the channels. Therefore, participants requested features to help set constraints on the data more effectively.

Further, participants also wanted to factor in economical (e.g., inflation, fed rate hike) and social (e.g., COVID, war, street protests) constraints into advanced analytics (P6, P9, P10, P15) since they had very significant impacts on businesses. Participants often dealt with these constraints for making decisions by turning to their intuition [97] and domain expertise. As a consequence, participants indicated that WIA techniques need to be enhanced to increase their reliability so that it can actively be used *"as an adjunct" (P8)* to inform their decisions. Therefore, future tools need to focus on feedback mechanisms between business professionals and advanced analytics which will allow professionals' expertise as well as external constraints to be easily integrated into predictions [36, 100].

# 8 DISCUSSION

Our two-part study confirms that business professionals are eager to independently conduct data-driven advanced analytics to inform their business decisions. Based on our findings, we discuss design recommendations for future business analytics systems to better support these advanced analytics needs.

## 8.1 Enhance Interpretation and Confidence

Our study reveals that business professionals are far from naive; they possess strong domain expertise and common sense, enabling them to critically assess predictions from automated tools. All participants indicated that WIA predictions cannot be followed blindly [47, 97], recognizing that even the most advanced models are based on limited data. Data often fails to capture all relevant constraints and domain-specific knowledge, resulting in predictions that may not be entirely accurate. Even commercial tools like Power BI [11], which provide advanced analytics features such as correlation insights or key influencers (akin to driver importance analysis), present these outputs to professionals but leave the responsibility of assessing their trustworthiness to the users. Participants emphasized the value of advanced analytics in complementing their decision-making and supporting their actions with data-driven evidence. However, they emphasized the need for systems that foster trust and confidence in the predictions they produce.

To achieve this, systems must go beyond merely displaying confidence levels buried within multiple features of commercial tools. They should transparently communicate prediction reliability and offer actionable steps to help professionals calibrate and validate these confidence levels effectively. Participants suggested that model accuracy scores or error metrics should be corroborated with risk factors that communicate potential negative outcomes, such as potential financial losses or the emergence of new competitors or products. Moreover, professionals desired guidance on next steps based on risk levels (e.g., a medium risk factor might suggest gathering more data and re-running the analysis). Several user studies [10, 36, 53, 100] have explored users' perceptions of trust and performance metrics in tasks such as model steering and predicting income group. Conducting similar studies specifically with business professionals for what-if analysis tasks can reveal other factors that enhance their trust in model predictions and, in turn, guide the design of advanced analytics systems that better meet their needs. Additionally, future research should investigate the requirements of technically proficient business professionals, focusing on analyses features that optimize their decision-making. This opens up a substantial opportunity for interdisciplinary collaboration to refine WIA techniques, making them more relevant and effective for business contexts.

## 8.2 Foster Communication and Team Collaboration

Participants consistently emphasized the importance of WIA in empowering business professionals to justify their decisions with data, providing them with a stronger voice when navigating organizational hierarchies. Since business decisions often require approval from executive stakeholders, it is essential for professionals to clearly communicate the rationale behind decision-making predictions generated by advanced analytics techniques. Additionally, this must be done with the understanding that both business professionals and executives usually lack expertise in data science or machine learning. This presents a clear opportunity to rethink traditional tools used.

For example, current dashboards like Tableau [90] or Power BI [11] typically display static textual summaries, such as the percentage of goals achieved or fixed drill-down paths into data segments [79, 92]. These high-level overviews, while useful, do not provide the flexibility needed for decision-making contexts [64]. Similarly, tables created in spreadsheet-based tools can be cumbersome and confusing [6], and cannot be modified on the fly during discussions or meetings. Further, translating insights from tables to individual presentations for every meeting requires substantial manual effort. By contrast, interactive visual tools accommodating advanced analytics could revolutionize these processes. Instead of static, one-way communication, such tools could allow professionals to dynamically explore various scenarios and their impacts on KPIs. For example, during a meeting, a business professional could adjust variables in real time to demonstrate how different marketing strategies may affect sales forecasts. These tools could facilitate live discussions and decision-making, as participants could see immediate outcomes of various strategies. In addition, these tools could be designed to quickly incorporate feedback from executives or other team members during the meeting itself [86], rather than requiring additional time to go back, rerun analyses and return with new findings. For instance, if an executive recalls a similar scenario from past experience or suggests adjusting a particular variable due to new constraints, the tool could instantly update the analysis to reflect that feedback. This kind of interactive, iterative process would not only save time but also foster collaboration, ensuring that decisions are informed by both data and experiential knowledge without the delays that are common in current workflows.

Further, emerging technologies such as natural language interfaces, interactive storytelling platforms [34, 87], and AI assistants [26, 67] must also be leveraged in such tools to go beyond merely presenting predictive results to also additionally explain, modify, and contextualize the predictions from advanced analytics.

## 8.3 Scale WIA with Decision Management Support

While WIA techniques allow for rapid computation of what-if scenarios and offer insights into various decision paths, participants observed that the number of scenarios and corresponding decision paths could grow exponentially [55, 56] as the number of driver variables and constraints increase. Hence, managing and reusing these analyses at scale presents a formidable challenge, especially as professionals increasingly gain access to real-time data. To prevent business professionals from becoming overwhelmed or, as one participant described, from *"getting lost in the sauce" (P9)*, it is essential to develop robust decision management systems that can effectively monitor, track, and log their analyses.

Additionally, such systems can also support professionals in formalizing their decision-making processes. For instance, they can assist in clearly identifying goals, drivers, and constraints, which is a critical step in structuring decisions and aligning them with business objectives. By providing this structure, future systems can be developed with targeted technological support to address the complex decision paths and associated challenges [20, 48]. For instance, these systems must be capable of handling large volumes of data and complex decision paths without compromising usability. Though such provenance-based systems have been widely developed for software developers [65, 81], data analysts [16, 58, 72], data scientists [52, 62, 85], and researchers [33, 41–43] for various analyses (e.g., visual analysis, statistical analysis, software development, etc.), there is gap in supporting domain experts like business professionals with accessible tools for using advanced analytics [20, 92]. These professionals require more user-friendly solutions tailored specifically for analyses that help inform data-driven decisions. For instance, professionals should be able to specify their hypotheses and parameters effortlessly without any need of technical expertise [6]. Hence, we develop our probe as an intuitive interactive interface requiring minimal external guidance. The tools should also support the seamless incorporation of domain knowledge and context, allowing professionals to leverage their expertise directly within the analysis, similar to how we allow the constraints to be specified. Furthermore, fostering trust in the models is essential, so tools must be designed to enhance transparency and make interpretations more accessible [23, 97, 99]. Finally, these solutions should simplify the management and comparison of multiple scenarios, enabling professionals to evaluate alternatives and make actionable decisions with greater ease and efficiency. By integrating formalization into decision management systems, these tools can bridge the gap between sophisticated analytics and the practical demands of business decision-making.

Further, decision management support systems can significantly enhance incorporation of WIA by enabling tools for ranking [32, 66], comparing, and recommending optimized decision paths [56]. Our study, along with others [19, 20, 92] indicate that such integration, especially as suggested by business professionals, will significantly improve their ability to leverage advanced analytics for truly data-driven decision-making. However, to maximize the success of these systems, it is essential to focus on usability and accessibility. For instance, neither programming-based tools nor systems supporting a multitude of models, features, complex visualizations, and workflows will meet the needs of business professionals. Beyond supporting WIA for structured tabular data, business use cases frequently involve qualitative data, such as expert opinions, contextual knowledge, or industry trends. Incorporating these unstructured data types into the decision-making process is also essential, as they provide critical context and insights that complement quantitative analyses. In this context, natural-language, query-based features powered by LLMs offer significant potential. These features have already demonstrated their utility in domains such as software testing [78] and cyber-physical systems [24], and could similarly enable business professionals to interact with complex advanced analytics and incorporation of textual data within it more intuitively. Therefore, besides the vast scope for machine learning modeling and optimization research, there is a huge opportunity for user experience and data and visual analytics communities to contribute to the development of scalable, user-friendly decision management systems that effectively support advanced analyses for decision-making.

## 8.4 Limitations

We acknowledge the limitations of our studies. First, our participant pool of business professionals across four departments has a limited sample size per department. Despite this, a similar number of participants have been studied in previous visualization and HCI research studies [1, 20, 45], and we chose few participants and covered multiple departments to ensure in-depth interviews and not have an extremely narrow collection of empirical data.

Second, our findings in the interview study are based on participants' self-reported experiences rather than performed workflows from their own use cases. Time constraints and the need to understand their advanced analysis techniques comprehensively led us to choose verbal articulation over demonstration. We addressed this by asking them to talk us through a specific but common marketing use case that was familiar to participants. Additionally, in the follow-up task-based study, participants' hands-on experience of using representative WIA techniques implemented in an interactive prototype was captured. We recognize that using such a prototype may introduce bias, where the specific visualization and interaction designs of the prototype can influence the results and the participants' reactions. Future work needs to tease apart these potential confounding factors. Further, we acknowledge that this study focused solely on comparing the proposed representative techniques to participants' current practices without evaluating them against other state-of-the-art tools. However, we believe that this step was essential to establish a baseline understanding of what business professionals perceive of advanced analytics techniques before introducing them to fully developed tools with extensive features, which is an exciting avenue for future work.

Further, we are aware that we used only one use case for both parts of the study. However, this was necessary to minimize participant variance, facilitate comprehension, and ensure participants' comfort in sharing their analyses without compromising sensitive information. For similar reasons, we used the marketing-specific use case even for non-marketing participants. To accommodate for this limitation, we provided examples of how the MMM task would be relevant to other departments, such as how MMM could enhance sales efficiency through optimized channel management and promotional strategies for sales managers. Additionally, we encouraged participants from non-marketing backgrounds to draw parallels between the techniques and their own use cases, enriching our findings with other contextualized examples.

# 9 CONCLUSION

In this paper, we conducted a two-part study with 22 business professionals from real-world enterprises. The first interview study aimed to understand the techniques, tools, and challenges in utilizing WIA for data-driven decision-making. Our findings show that business professionals perform a range of advanced analyses without relying on data analysts due to various logistical constraints. However, they often use rudimentary methods that are insufficient for effective decision-making. We conduct a follow-up task-based study with the same participants to elicit business professionals' feedback on the benefits and opportunities of enhancing advanced analysis techniques. In this follow-up study, the participants used four representative WIA techniques identified in the first interview study, which we implemented in a visual analytics prototype as a probe for them to use hands-on to make data-driven decisions. Our findings show that these techniques improved decision-making speed and confidence while also highlighting the need for better dataset preparation, risk assessment, and domain knowledge integration. Finally, we make design recommendations for future business analytics systems based on these insights.

# ACKNOWLEDGMENTS

We thank Sigma Computing Inc., where this work was initiated as an internship project. We also thank Joe Hellerstein, Arjun Srinivasan, and the anonymous reviewers for their helpful feedback on the paper draft.